\documentclass[12pt,preprint]{aastex}

\usepackage{graphicx,epsfig,natbib,color}

\def \<{\langle}
\def \>{\rangle}

\begin{document}
\title{Pseudo-Dipole Signal Removal from WMAP Data}

\author{Hao Liu\altaffilmark{1} and Ti-Pei Li\altaffilmark{1,2,3}}

\affil{$^1$ Key Laboratory of Particle Astrophysics, Institute of High Energy Physics,
Chinese Academy of Sciences, Beijing, China  {\tt (liuhao@ihep.ac.cn)}}

\affil{$^2$ Department of Physics \& Center for Astrophysics, Tsinghua University, Beijing, China {\tt(litp@tsinghua.edu.cn)}}

\affil{$^3$ Department of Engineering Physics \& Center for Astrophysics, Tsinghua University, Beijing, China}

\begin{abstract}
It is discovered in our previous work that different observational systematics, e.g., errors of antenna pointing directions, asynchronous between the attitude and science data, can generate pseudo-dipole signal in full-sky maps of the cosmic microwave background (CMB) anisotropy published by The Wilkinson Microwave Anisotropy Probe (WMAP) team. Now the antenna sidelobe response to the Doppler signal is found to be able to produce similar effect as well. In this work, independent to the sources, we uniformly model the pseudo-dipole signal and remove it from published WMAP7 CMB maps by model fitting. The result demonstrates that most of the released WMAP CMB quadrupole is artificial.
\end{abstract}


\section{PSEUDO-DIPOLE SIGNAL}
The WMAP mission makes measurements of temperature with two antennas, A and B, and record in time-order the difference between the two antenna temperatures, $T_A-T_B$, which is called the time-order data (TOD). The observed CMB signal is contaminated by Doppler effect induced by the joint motion of the solar system and the spacecraft. The aroused dipole difference signal can be calculated by
\begin{equation}\label{equ:dp1}
d=\frac{T_0}{c}~\bf v\cdot (\bf n_{_A}-\bf n_{_B})\,,
\end{equation}
where $T_0=2.725$\,K is the CMB monopole, $c$ is the speed of light, $\bf{v}$ is the joint velocity, $\bf{n}_{_A}$ and $\bf{n}_{_B}$ are the unit direction vectors of the antenna A and B respectively ~\citep{hin09}. In this paper the vectors $\bf v$, $\bf n_{_A}$, and $\bf n_{_B}$ are defined in the spacecraft coordinate system. The errors of line-of-sight (LOS) vector, $\Delta \bf{n}_{_A}$ and $\Delta \bf{n}_{_B}$, will produce a pseudo-dipole difference signal
\begin{eqnarray}\label{equ:pds}
d^* &=& \frac{T_0}{c}~\bf{v}\cdot (\rm\Delta\bf{n}_{_A}-\rm\Delta\bf{n}_{_B}) \nonumber \\
&=&\frac{T_0}{c}~\bf{v}\cdot \rm\Delta\bf{n}\,.
\end{eqnarray}

The dipoles for each observations have to be removed from the raw data before map-making because their intensities are roughly 10 to 20 times greater than those of the CMB anisotropies. A small error of antenna direction will produce an error in predicted dipole intensity and then cause a pseudo-dipole signal in the resulting CMB map more noticeably through the Doppler dipole subtraction process\footnote{A true and pure time-order dipole signal produces only a dipole component in the final CMB map, but a time-order pseudo-dipole signal can produce more components, especially quadrupole, on the final CMB map. See Fig.~1.} .  For example, a LOS error of $\sim7'$, just about a half-pixel in the WMAP resolution, can consequently cause the dipole signal to be deviated by 10--20\,$\mu$K, which can not be ignored compared to the very weak CMB signal. An asynchronous between the attitude and differential data can also produce the pseudo-dipole signal. The WMAP mission uses two separate clocks for the attitude data and science data respectively. Therefore, if there is a small constant timing error, there will be a constant direction difference between the "observed pixel" and the true pixel. This has the same effect as a constant LOS error in spacecraft coordinates \citep{liu10}.

\section{SIDELOBE PICKUP}
Another possible source of pseudo-dipole signal in released WMAP maps is the sidelobe signal contamination. Like all radio telescopes, the WMAP antennas have both main beam response and sidelobe response. The WMAP antenna sidelobe response was described by \citet{barnes03} and the corresponding data file is publicly available\footnote{The data file of WMAP sidelobe response can be found at http://lambda.gsfc.nasa.gov/product/map/dr4/farsidelobe\_info.cfm.}. The data files are fits format full sky maps in spacecraft coordinates in which the sidelobe responses are given in normalized gain $G$, where the normalization rule is that the summation of all gains for one antenna (including the main beam) equals to $N$, the number of pixels in the map. Thus for each differential observation, the recorded difference signal is $\sum_{i=0}^{N-1} (G_i^A-G_i^B)T_i/N$. Let $k$ denotes pixels corresponding to a certain sidelobe, then the overall sidelobe response is $\sum_k(G_k^A-G_k^B)T_k/N$. The overall sidelobe response should contribute a pseudo component to the differential Doppler signal
\begin{equation}\label{equ:sl}
d_{sidelobe}^*=\frac {T_0}{c}\,\bf{v}\cdot\rm\it\sum_k\frac{(G_k^A-G_k^B)\, \bf n_{_k}} {N}\,.
\end{equation}
The spacecraft LOS vectors and $\bf{n}_{_k}$ are all constant vectors in spacecraft coordinates. Since $G$ are normalized gains, it can also be treated as constant. Therefore we have two constant vectors: $\Delta\bf{n}_{_A}^*=\rm\it\sum_k G_k^A\, \bf{n}_{_k}/\it{N}$ and $\Delta\bf{n}_{_B}^*=\rm\it\sum_k G_k^B\, \bf{n}_{_k}/\it{N}$. Since the sidelobe gain is much smaller than the main lobe gain, $\Delta\bf{n}_{_A}^*$ and $\Delta\bf{n}_{_B}^*$ are two small constant vectors, and
\begin{eqnarray}\label{equ:dp_sl}
d_{sidelobe}^* &=& \frac{T_0}{c}\,\bf{v} \cdot (\rm\Delta\bf{n}_{_A}^* -\rm\Delta\bf{n}_{_B}^*) \nonumber \\&=& \frac{T_0}{c}\,\bf{v} \cdot \rm\Delta\bf{n}^*\,.
\end{eqnarray}
By comparing Eq.~2 and Eq.~4, we can see that the entire sidelobe Doppler pickup can be exactly described by a small constant deviation to the differential spacecraft LOS vector. Similarly, the effect of  the sidelobe response uncertainty on the differential Doppler dipole is also equivalent to introducing a small differential LOS error  $\Delta\bf{n}^*$. It is important to notice that, even if the LOS is absolutely accurate, this pseudo-dipole signal from the sidelobe response uncertainty can still exist.

The amplitude of the equivalent LOS error induced by sidelobe Doppler pickup can be estimated by $\arcsin(|\Delta\bf{n}_{_A}^*|/|\bf{n}_{_A}|)$, which are $\sim 50' - 75'$ for different bands. The WMAP team believes that the Galactic sidelobe pickup is neglectable except for the K-band, the reason is largely that the Galactic emission is much weaker in other bands than in the K-band ~\citep{barnes03}. The case is significantly different for the Doppler signals because they have almost the same amplitude in all bands. Moreover, the Doppler signal is as strong as several mK all over the sky; however, the Galactic emission is strong only for the low Galactic latitude and the strength decreases rapidly for higher latitude. Thus the overall power of the Doppler signal is about $\sim 10-40$ times higher than the overall Galactic foreground emission power, and the effect of sidelobe Doppler signal pickup is at least 10 times stronger than the sidelobe Galactic pickup.

The WMAP antenna sidelobe gain patterns are estimated by ground-based measurements and in-flight lunar measurements. For the K-band,  the WMAP first-year in-flight gain measurements are $60\%$ systematically brighter than ground-based measurements. The WMAP team scaled up the ground-based measurements by $30\%$ and scaled down the lunar results by $30\%$ to yield a best-guess sidelobe gain map with an overall calibration uncertainty $\sim 30\%$~\citep{barnes03}, thus the amplitude of the equivalent LOS error induced by the sidelobe Doppler pickup uncertainty could be up to $\sim 15' - 22'$ (estimated by $\arcsin(|\Delta\bf{n}_{_A}|/|\bf{n}_{_A}|)\times\rm30\%$), which is much more than enough to produce considerable pseudo-dipole contamination.  From the WMAP releases we find that the WMAP7 sidelobe file is exactly the same with WMAP1 for the K-band, and for other bands  the sidelobe files of a single survey year are just  slightly changed from year-1 to year-7, indicating that their uncertainties have not been essentially improved and can not be ignored.

\section{MODELING AND REMOVING PSEUDO-DIPOLE SIGNAL}
The approach we use to remove the pseudo-dipole signal is same to the WMAP foreground removal.  Different effects arising foreground are separately modeled with template maps: the 94\,GHz dust map $t_d$ for dust emission~\citep{Fink_dust_99}, the full-sky H$\alpha$ map $t_{_H}$  for free-free emission~\citep{Fink_ff_03}, and the synchrotron template $t_s$ derived from K, Ka bands for synchrotron emission~\citep{ben03, Gold09_fg}. The foreground emission is removed by the WMAP team~\citep{ben03, Gold09_fg} with model fitting to get the clean temperature
\begin{equation}\label{equ:fg}
t_{clean}=t'-(c_d\, t_d+c_{_H}\, t_{_H} +c_s\, t_s)\,,
\end{equation}
where the coefficients $c_d$, $c_{_H}$, $c_S$ are determined by weighted fitting to make $t_{clean}^2=\min$.

\begin{figure}[t]
\begin{center}
\includegraphics[height=5cm, angle=90]{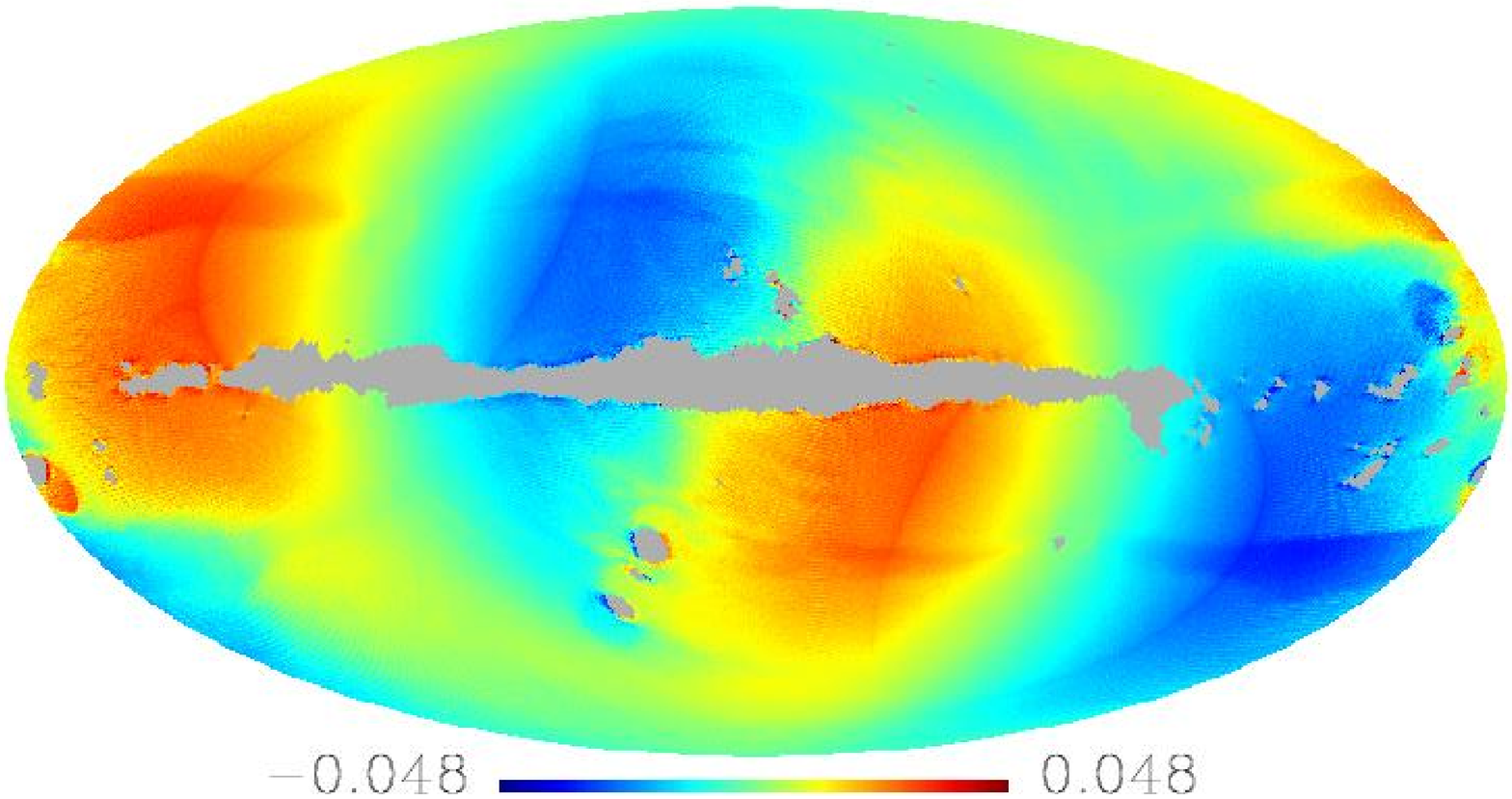}
\includegraphics[height=5cm, angle=90]{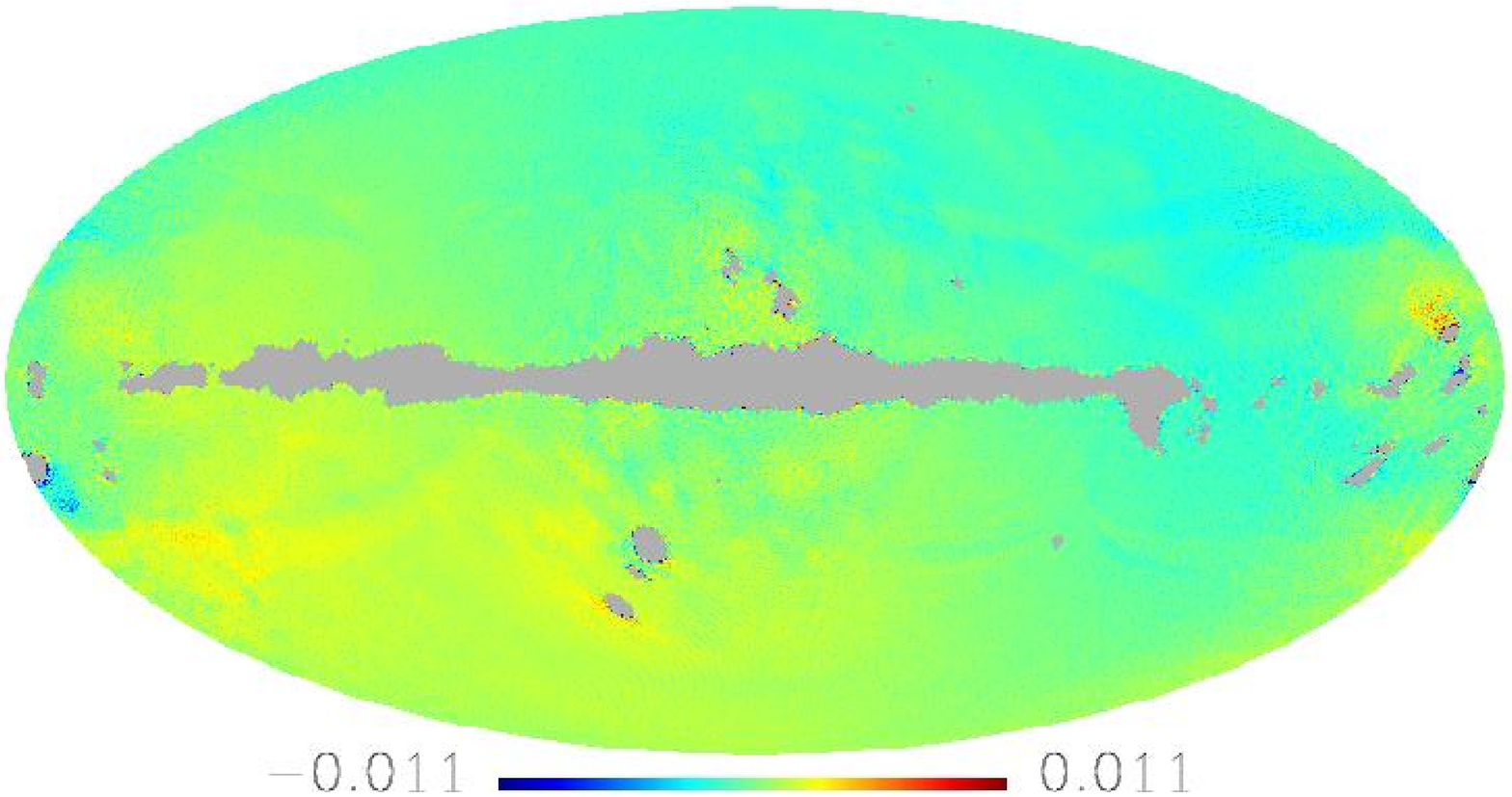}
\includegraphics[height=5cm, angle=90]{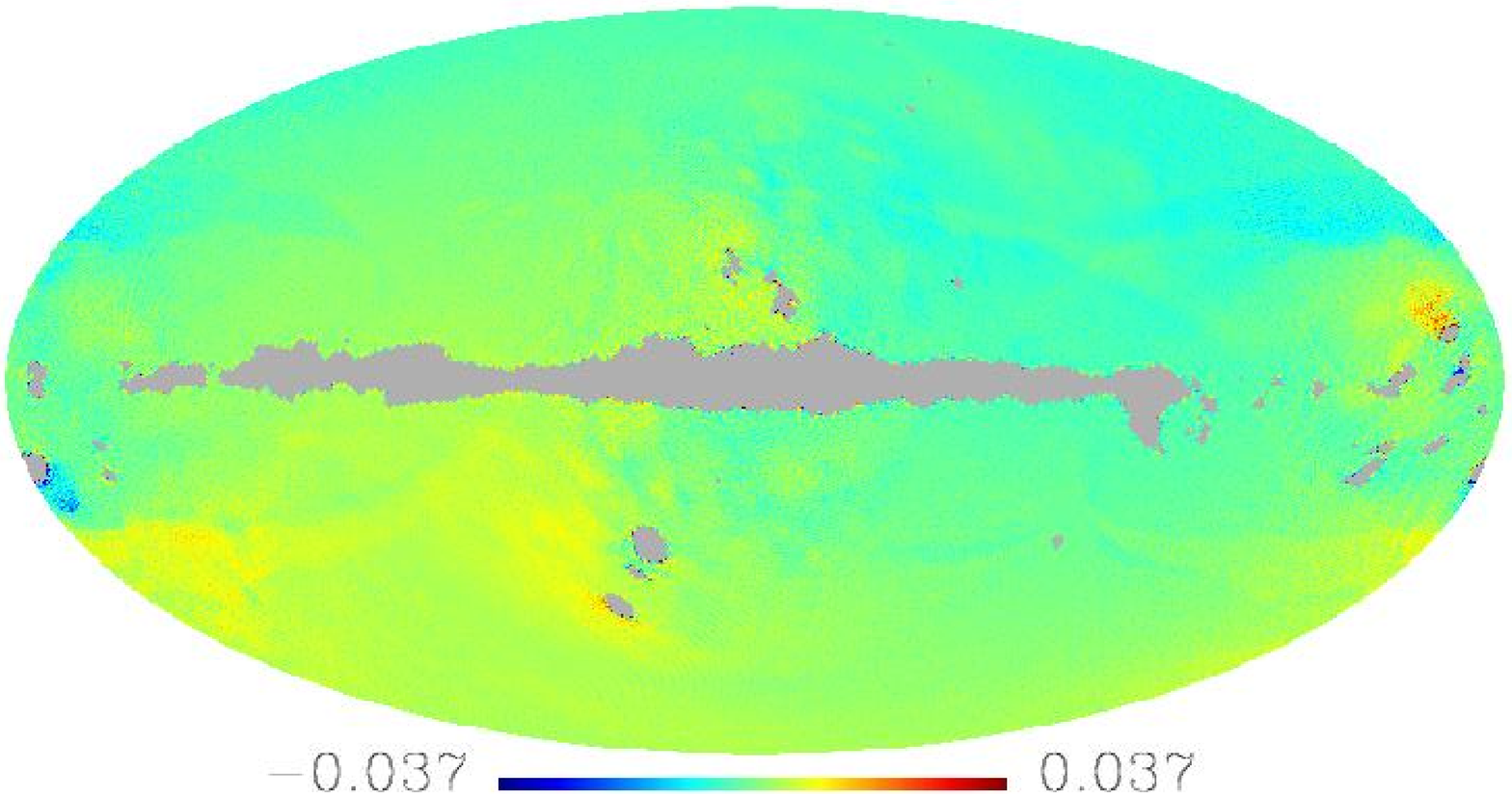}
\end{center}
\caption{ \footnotesize{Induced map from the pseudo-dipole signal produced by a differential  direction error $\Delta \bf{n}$, for WMAP7 year-1 Q1-band, in mK and in Galactic coordinates. From left to right: $\delta t_x$ with $\Delta \bf{n}\rm=(0.01, 0, 0)$ , $\delta t_y $ with $\Delta \bf{n}\rm =(0, 0.01, 0)$, $\delta t_z$ with $\Delta \bf{n}\rm=(0, 0, 0.01).$ Results for other years and other bands are similar to this. }}
\end{figure}

The pseudo-dipole signal in WMAP temperature maps can be removed with the same approach for foreground contamination. The overall effect of the three systematics mentioned in \S1 and \S2, LOS error, time drift, and sidelobe Doppler pickup, can be generally described by an equivalent deviation $\Delta\bf{n}$ of the differential antenna direction. The deviation induced by the pseudo-dipole signal in a released WMAP map can be easily modeled. For each measured temperature difference in the TOD used to produce the map, we substitute the pseudo-dipole signal calculated by Eq.~2 for an assumed $\Delta\bf{n}$ to produce a new TOD\,\footnote{In calculation, after adding a deviation to the LOS vector, we always re-scale the LOS vector to keep it unitary.} . The temperature map produced by map making\,\footnote{Our map-making codes are publicly released on the website of Tsinghua Center for Astrophysics at http://dpc.aire.org.cn/data/wmap/09072731/release\_v1/source\_code/v1/ and on the CosmoCoffee Forum at http://cosmocoffee.info/viewtopic.php?p=4525\#4525.} from the new TOD can be used as a model of the pseudo-dipole signal in the released map. In calculation we use the spacecraft coordinate system $(X, Y, Z)$, where the $X$ axis is parallel to plane of radiators, $-Z$ is the anti-sun direction of the spin axis, and $Y$ is perpendicular to both. The WMAP spacecraft scans the sky with a hybrid motion mode consists of rotation and precessing. In spacecraft coordinates, the LOS unit vectors of its two antennas are close to $(x,y,z)=(0, 0.94, -0.33)$ and $(x,y,z)=(0, -0.94, -0.33)$, and the spacecraft rotation is around the $Z$-axis. Suppose the overall LOS error $\Delta\bf{n}$ from all such effects is made up of three small vectors $(\delta, 0, 0)$, $(0,\delta, 0)$ and $(0, 0, \delta)$ with $\delta=0.01$, each alone on the final map induces its own full-sky distribution of deviation $\delta t_x$, $\delta t_y$ and $\delta t_z$, respectively. The induced deviation upon the released WMAP7 year-1 Q1-band map, $\delta t_x$, $\delta t_y$, and $\delta t_z$ are shown in Fig.~1. Results for other years and other bands are similar.

It is easy to see from Fig.~1 that $\delta t_y$ and $\delta t_z$ are highly correlated. In order to avoid degeneracy issue we only use $\delta t_x$ and $\delta t_y$ in model fitting. The clean full sky temperature map
\begin{equation}\label{equ:real t}
t_{clean}=t'-(c_x\, \delta t_x+c_y\, \delta t_y)\,,
\end{equation}
where $t'$ is the corresponding WMAP CMB temperature map and the coefficients $c_x$ and $c_y$ can be determined by minimizing $t_{clean}^2$. Using the standard IDL program "regress"  we get $c_x=-0.35$, $c_y=-0.78$ for the WMAP7 year-1 Q1-band map. We also model and remove the pseudo-dipole signal from the released WMAP7 year-1 to year-7 maps of Q1, Q2, V1, V2, W1, W2, W3 and W4 bands, separately\footnote{For example, if the two templates $\delta t_x$ and $\delta t_y$ are derived from WMAP7 year-1 Q1 band, then they are used to fit the WMAP7 year-1 Q1 band single year CMB temperature map. The WMAP7 single year CMB temperature maps can be found at http://lambda.gsfc.nasa.gov/product/map/dr4/maps\_forered\_da\_r9\_i\_1yr\_get.cfm.}.  From the clean maps we calculate their power spectra and residual quadruples. Table 1 lists the obtained residual quadrupoles of different bands. The overall average clean quadrupole power for all bands is found to be $\sim 17.1 \rm{\mu K}^2$, only $\sim14\%$ of what released by the WMAP team\footnote{The WMAP pseudo-Cl power spectrum for the quadrupole can be found in the WMAP1 release ($123\, \rm{\mu K}^2$ for quadrupole) at http://lambda.gsfc.nasa.gov/data/map/powspec/wmap\_binned\_tt\_powspec\_yr1\_v1p1.txt. Although the WMAP team never gave a pseudo-Cl quadrupole in later releases, we have tested and made sure that this value is almost the same in all releases from WMAP1 to WMAP7.}, indicating that most of the published WMAP CMB quadrupole is artificial.

\begin{table}[t]{Table 1: The residual quadrupole power  ($\mu$K$^2$)}\\[1ex]
\begin{tabular}{ccccccccc}  \hline
Wave Band  &  Q1 & Q2 & V1 & V2 & W1 & W2 & W3 & W4 \\
Residual Quadrupole& 16.0 & 13.6 & 10.4 & 13.7 & 27.5 & 20.2 & 12.6 & 23.3\\
Standard Deviation& 7.0 & 3.6 & 3.9 & 2.8 & 7.0 & 8.5 & 5.3 & 27.4 \\
\hline
\end{tabular}
\tablecomments{For each band, the average residual quadrupole (line 2) and the sample standard deviation (line 3) is calculated from the 7 single year results.}
\end{table}

\section{DISCUSSION}
We have found that at least three systematical effects inducing pseudo-dipole signal in WMAP raw differential temperature data can be summarized into a uniform description: the equivalent differential direction error $\Delta\bf{n}$. There could be more sources with similar attributes to these three, but they can all be covered equivalently by a simple $\Delta\bf{n}$. For example, the LOS is determined by observing the Jupiter; however, for each observation, the relative direction of Jupiter provides no more than three conditions: $(x,y,z)$, but we need four conditions to definitely determine the attitude of the spacecraft: $(x,y,z)$ and rotation. What's more, the LOS is determined together with the antenna main beam response; therefore, the uncertainty of the beam response is also mixed into the LOS uncertainty.
   Moreover, even the WMAP team themselves have found that some systematical effect can produce quadrupole-like deviations on the final CMB map (Jarosik et al. 2007, Fig. 3).
Recently, \citet{rou10} pointed out that a small timing error during the step of calibration of the raw TOD could correspond to adding a pseudo-dipole difference signal with no effect on positional data.
It is impossible to exactly estimate all such effects, especially when some of them could be unknown for now, but it is easy to model the impact of $\Delta\bf{n}$ upon the final CMB map and remove it by model fitting. This is apparently much more feasible and effective than estimating and removing all possible sources one by one.

  What is important for model fitting is the feature of a template map, not the absolute amplitude. For example, the WMAP foreground removal will be the same even if the three foreground templates $t_d$, $t_{_H}$ and  $t_s$ are all doubled. In this work, the result of removing pseudo-dipole signal is not dependent to what amplitude is assumed for the direction error: larger amplitude leads to lower fitted coefficient in model fitting and the final correction to the CMB map is the same. Here it is worth to mention that the characteristic feature of template map of the pseudo-dipole signal induced by a pointing or timing error obtained from our previous work \citep{liu10} or shown in Fig.~1 of this work  has already been  confirmed by other independent works \citep{mos10,rou10}.

Our result demonstrates  that a pseudo-dipole field can produce deviations in the sky temperature map with a structure very similar to the CMB quadrupole pattern published by the WMAP team. The template map of pseudo-dipole induced temperature deviation shown in Fig.~1 are generated from a differential dipole field which is completely determined by the spacecraft velocity and overall-equivalent error of differential direction without using any CMB signal.  The pseudo-dipole signal in the differential datum for a WMAP observation distorts the temperatures for corresponding sky pixels varying observation by observation.  Therefore, the pseudo-dipole induced temperature map is generated by the pseudo-dipole signal combined with the WMAP scan pattern,  which is highly relative to the ecliptic plane~\citep{hin07, li09}.   The WMAP CMB quadrupole component being highly aligned and close to the ecliptic plane is a long time puzzle in cosmology, which now can be naturally explained by the pseudo-dipole signal effect.  Keeping insist that the published WMAP quadrupole is really cosmological origin now becomes more difficult:  it is needed the primordial density fluctuations not only occasionally being laid down in the plane of the solar system, but also occasionally having almost the same phase as the scan strategy of the WMAP mission!

\acknowledgements
This work is supported by the National Natural Science Foundation (Grant No. 10821061), National Basic Research Program of China (Grant No. 2009CB-824800), the CAS project KJCX2-YW-T03 and China Postdoctoral Science Foundation funded project H91I21734A. The data analysis made use of the WMAP data archive and the HEALPix software package~\citep{gor05}.

\bibliographystyle{apj}

\end{document}